\documentclass[conference]{IEEEtran}
\usepackage{epsfig}
\usepackage{graphicx}
\usepackage[ansinew]{inputenc}
\usepackage{amsmath}
\usepackage{cite}
\usepackage{pgfplots}
\pgfplotsset{compat=newest}
\usetikzlibrary{plotmarks}
\usepgfplotslibrary{patchplots}
\usepackage{grffile}
\usepackage[letterpaper, left=1in, right=1in, bottom=1in, top=0.75in]{geometry}

\makeatletter
\renewcommand*{\env@matrix}[1][*\c@MaxMatrixCols c]{%
  \hskip -\arraycolsep
  \let\@ifnextchar\new@ifnextchar
  \array{#1}}
\makeatother

\begin{document}

\title{On Partly Overloaded Spreading Sequences with Variable Spreading Factor}
\author{\IEEEauthorblockN{Michael Karrenbauer, Andreas Weinand, and Hans D. Schotten\\}
\IEEEauthorblockA{Chair for Wireless Communications and Navigation\\
University of Kaiserslautern, Germany\\
Email: \{karrenbauer,weinand,schotten\}@eit.uni-kl.de}
}\maketitle

\begin{abstract}
Future wireless communications systems are expected to support multi-service operation, i.e. especially multi-rate as well as multi-level quality of service (QoS) requirements. This evolution is mainly driven by the success of the Internet of Things (IoT) and the growing presence of machine type communication (MTC).
Whereas in the last years information in wireless communication systems was mainly generated or at least requested by humans and was also processed by humans, we can now see a paradigm shift since so-called machine type communication is gaining growing importance. Along with these changes we also encounter changes regarding the quality of service requirements, data rate requirements, latency constraints, different duty cycles et cetera. The challenge for new communication systems will therefore be to enable different user types and their different requirements efficiently.
In this paper, we present partly overloaded spreading sequences, i.e. sequences which are globally orthogonal and sequences which interfere with a subset of sequences while being orthogonal to the globally orthogonal sequences. Additionally, we are able to vary the spreading factor of these sequences, which allows us to flexibly assign appropriate sequences to different service types or user types respectively. We propose the use of these sequences for a CDMA channel access method which is able to flexibly support different traffic types.
\end{abstract}
\section{Introduction}
\begin{figure}%[ht]
	\centering
		\includegraphics[width=0.8\columnwidth]{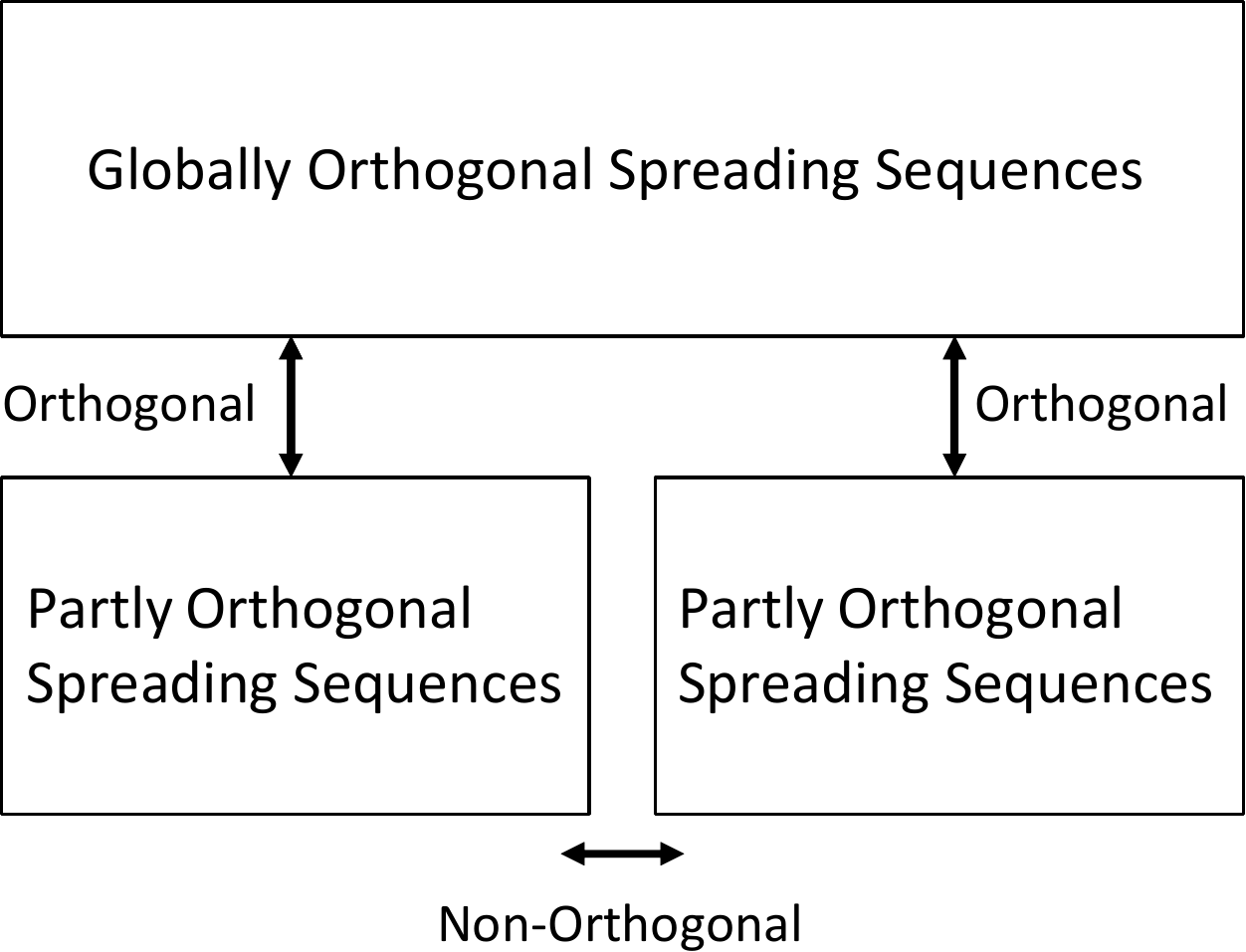}
	\caption{Partition of spreading sequences}
	\label{fig:codes}
\end{figure}

The ability to efficiently support different traffic types with different requirements is a crucial one for today's wireless systems and is expected to gain even more importance in the future, along with but not limited to the success of the Internet of Things (IoT) and the proceeding of industrial communication technologies. One specific example of a system, which would benefit of supporting different traffic types simultaneously and which we would like to address, would be a wireless communication system in an industrial environment. In today's production sites, wireless communication systems are serving either human purposes (in the following referred to as ``best-effort traffic'') or non-human purposes (in the following referred to as ``machine-type traffic'') exclusively, because a flexible support of both traffic types is not possible, which leads to a waste of spectrum.
In accordance with what intuition would tell us, analysis of machine-to-machine (M2M) traffic reveals that it has got specific features which hold true for most cases \cite{Nikaein2013}. In table \ref{tab:traffic}, the different message characteristics for best-effort and for machine-type traffic are shown.
\begin{table}%[h]
\caption{Characteristics of best-effort and machine-type traffic}
\centering
\begin{tabular}{l|l|l}
\textbf{Parameter} 						 & \textbf{best-effort traffic}  		& \textbf{machine-type traffic}\\
\hline
packet length									 & long															& short\\
number of packets						   & low  														& high\\
duty cycle										 & high															& low\\
traffic direction							 & downlink-												& uplink-\\
															 & dominant													& dominant\\
latency constraints						 & medium														& medium up to\\
															 &																	& real-time\\
traffic generation						 & human triggered									& periodic or\\
															 &																	& event-driven\\
Bit Error Rate								 & medium														& high up to\\
requirements 									 &      														& ultra-reliable
						
\end{tabular}
\label{tab:traffic}
\end{table}
Our idea, that we would like to present in the following sections, is to use a CDMA-based MAC which possesses two subsets of spreading sequences, as depicted in figure \ref{fig:codes}, which can be assigned to a corresponding traffic type. The first subset of sequences is consisting of sequences which are mutually orthogonal to every other sequence in this subset and in ever other subset of sequences. The second subset of sequences is consisting of sequences which are mutually orthogonal to every sequence in the first subset but interfere with sequences in the second subset of sequences. Intuition suggests to assign sequences of the first group of sequences to traffic with high bit error rate requirements, i.e. machine-type traffic, while assigning sequences of the second group of sequences to traffic with lower reliability requirements, i.e. best-effort traffic. 
In a second step, we would like to combine this idea with a variable spreading factor to gain additional flexibility. Assuming a constant frame length, we then have a total of three degrees of freedom: the FEC coding rate, the spreading factor and the amount of overloading in the CDMA system. %Our goal is to set these parameters optimally for different traffic types and traffic models respectively.
The rest of this paper is organized as follows: in the following section, an overview of related work is given. In section III, the generation of our proposed spreading sequences is explained. Section IV explains the support of variable spreading factors within the proposed CDMA system. In section V, numerical evaluations using computer simulations are performed. A conclusion is drawn in section VI. 

\section{Related Work}
The CDMA MAC we are proposing in this paper is related to different topics already known in the literature which we would like to shortly mention in the following.
\subsubsection{Hadamard Matrices}
The spreading sequences used in this paper are based on a well-known family of orthogonal spreading sequences called Hadamard sequences \cite{Sylvester1867}\cite{Hadamard1893}. A family $F=\{f_i(t)|i=1,2...N\}$ of $N$ binary sequences is called orthogonal, if any two sequences $x(v)$ and $y(v)$ of length $N=2^n$ are mutually orthogonal, i.e. their synchronous cross-correlation 
\begin{equation}
C_{x,y} = \sum_{v=0}^{N-1}{x(v)\cdot y(v)},
\end{equation}
is zero.

A matrix $H$ consisting of $N$ different sequences of length $N$, forming an $N \times N$ matrix is called a Hadamard matrix, if the following equation applies: 
\begin{equation}
H \cdot H^T=H^T \cdot H=N \cdot I
\end{equation}
with $I$ being the identity matrix and $H^T$ being the transposed matrix $H$.
%----------------------------------------------------------------------------------------------------------------------------------------------
\subsubsection{Orthogonal Variable Spreading Factor}
\begin{figure}[ht]
	\centering
		\includegraphics[width=\columnwidth]{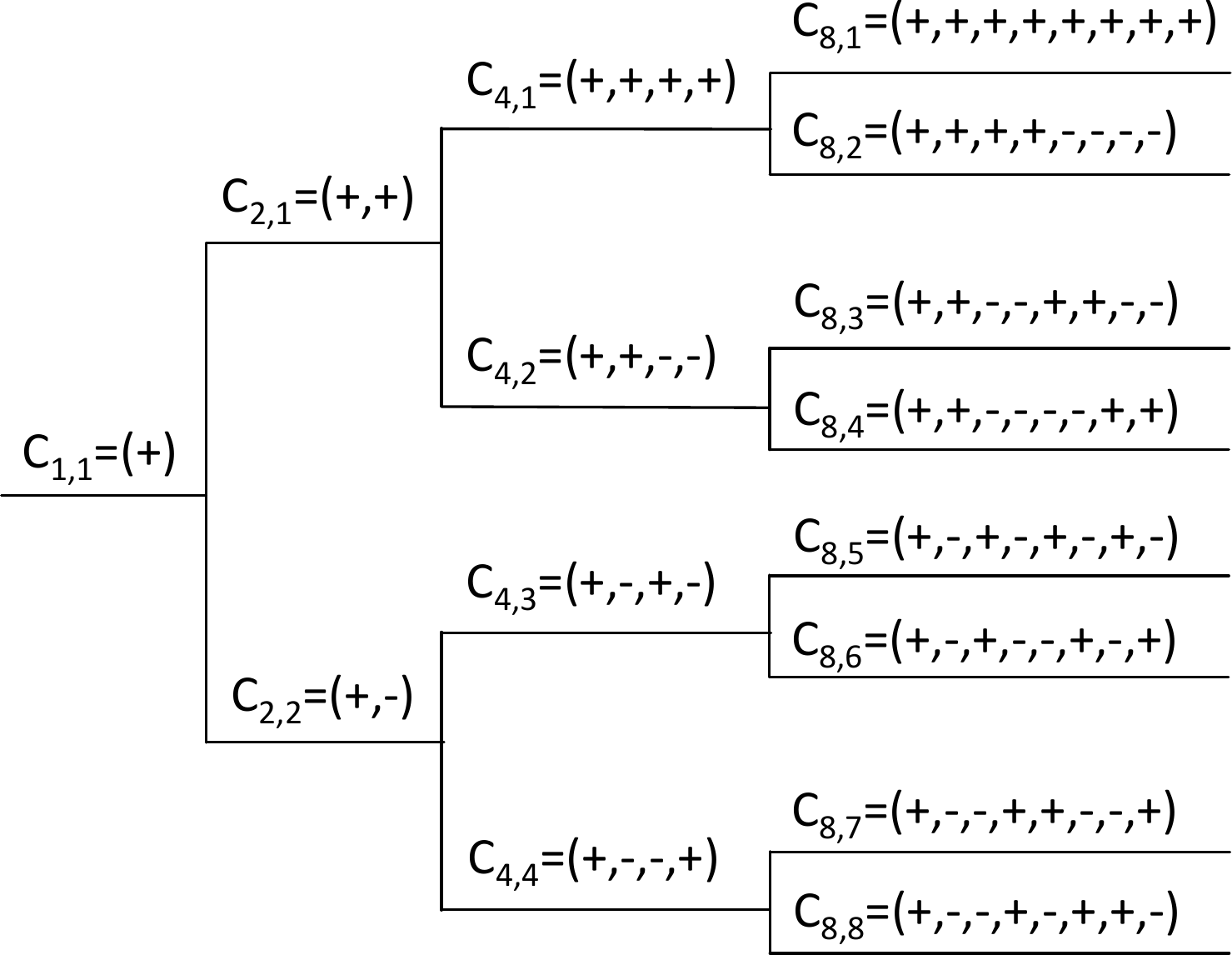}
	\caption{Orthogonal Variable Spreading Factor (OVSF) code tree}
	\label{fig:ovsf}
\end{figure}

To support multi-rate operation within a CDMA system, the orthogonal variable spreading factor (OVSF) code has been introduced in the literature \cite{Adachi1997}. The idea is to use Hadamard codes of different lengths and to organize them in a tree structure according to figure \ref{fig:ovsf}.
The generation of the sequences differs slightly from Sylvester's method \cite{Sylvester1867} and is defined recursively leading to the following sequence of matrices:
\begin{equation}
H_1 = \begin{pmatrix}[r]
     1
\end{pmatrix},
H_{2^{n}} = \begin{pmatrix}[l]
     H_{2^{n-1}}\otimes(1,1)\\
     H_{2^{n-1}}\otimes(1,-1)
\end{pmatrix},
\end{equation}
with $\otimes$ being the Kronecker product. As we can easily check using equation 2, the generated matrices are Hadamard matrices.
As a consequence, codes of the same layer are orthogonal. Additionally, any two codes of different layers are also orthogonal as long as one of them is not a mother code of the other.
%-----------------------------------------------------------------------------------------------------------------------------------------------
\subsubsection{Overloaded CDMA}
As we could see, an orthogonal set of Hadamard codes of length $N$ consists of $N$ different sequences which are mutually orthogonal. If we are in need of more than $N$ different spreading sequences and are not intending to increase the code length, i.e. the spreading factor (SF), we have to add sequences which are not mutually orthogonal to every other sequence of the Hadamard set and thus bring additional interference with them. This technique is called overloaded CDMA and is well known in the literature \cite{Hosseini2011}.
Following the notation of \cite{Yang2000} and \cite{Amadei2002}, the maximum cross-correlation between an additional sequence $f$ of length $N$ and every sequence $h_i$ in the Hadamard code set $H$ is defined as
\begin{equation}
C_f=\displaystyle\max_{i} C_{f, (h_i+i_0)}
\end{equation}
where $i_0$ is the all-zero or the all-one sequence of length $N$. Furthermore, the authors of \cite{Yang2000} and \cite{Amadei2002} also stated the minimum achievable correlation value for all sequences f, $C_{min}(N)=\displaystyle\min_{f}C_f$ which is given by
\begin{equation}
C_{min}(N)=2^{n/2}=\sqrt{N}
\end{equation}
for any even integer n.
Additional sequences fulfilling equation (5) and thus minimizing additional interference are referred to as ``Quasi-orthogonal sequences'' in the literature. In \cite{Yang2000}, quasi-orthogonal sequences for the use in CDMA systems are described. The authors of \cite{Amadei2002} combine this approach with a variable spreading factor and use these sequences in a multi-carrier CDMA system supporting multiple classes of users with respect to different data rates only. To the best of the authors knowledge, a CDMA system using only partly overloaded spreading sequences and supporting multiple QoS levels is not yet published in the literature.
%-----------------------------------------------------------------------------------------------------------------------------------------------
\subsubsection{Trade-Off between CDMA and FEC}
Improving signal robustness by adding additional information can either be done by applying channel coding or by increasing the spreading factor. This trade-off is investigated for a multi-carrier CDMA system in\cite{Kaiser1996}.
\section{Partly Overloaded Spreading Sequences}
\begin{figure}%[ht]
	\centering
		\includegraphics[width=\columnwidth]{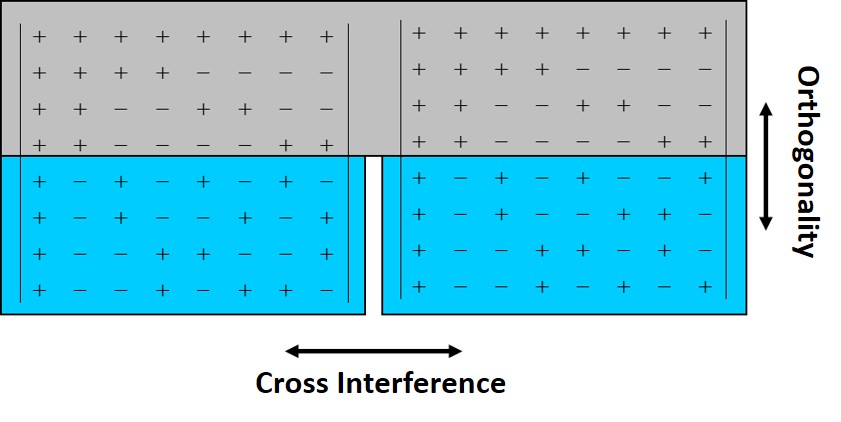}
	\caption{Overloaded Hadamard set}
	\label{fig:QOSS}
\end{figure}

For our approach, we generate the spreading sequences as follows. First of all we start with the generation of spreading sequences with $SF=2^n$ according to equation (3), which leads to a Hadamard matrix. On the left-hand side of figure \ref{fig:QOSS} the corresponding Hadamard matrix of $SF=8$ is shown as an example. We are considering every row of this matrix as being one spreading sequence. As it is known in the literature \cite{Koukouvinos2008}, we are allowed to interchange columns (as well as rows) of a Hadamard matrix which will lead to another Hadamard matrix of $SF=8$. Since, as we can observe, neighboring columns of the above half of our Hadamard matrix are equal, this subset of spreading sequences does not change, when we interchange these columns. In the below subset however, we create additional spreading sequences by interchanging these columns. Using equation (2), it can be shown, that this new matrix, consisting of the original Hadamard matrix with interchanged columns is also a Hadamard matrix. The new spreading sequences are of course also orthogonal to the unchanged above part of the spreading matrix, but bring additional interference with the original below part of the matrix. %Using equations (4) and (5) it can be shown, that this additional interference is minimal, which makes these sequences ``Quasi-orthogonal sequences''.
A code set consisting of sequences of length 8 is depicted in figure \ref{fig:QOSS}.
\section{Partly Overloaded Spreading Sequences with Variable Spreading Factor}
\begin{figure}%[ht]
	\centering
		\includegraphics[width=\columnwidth]{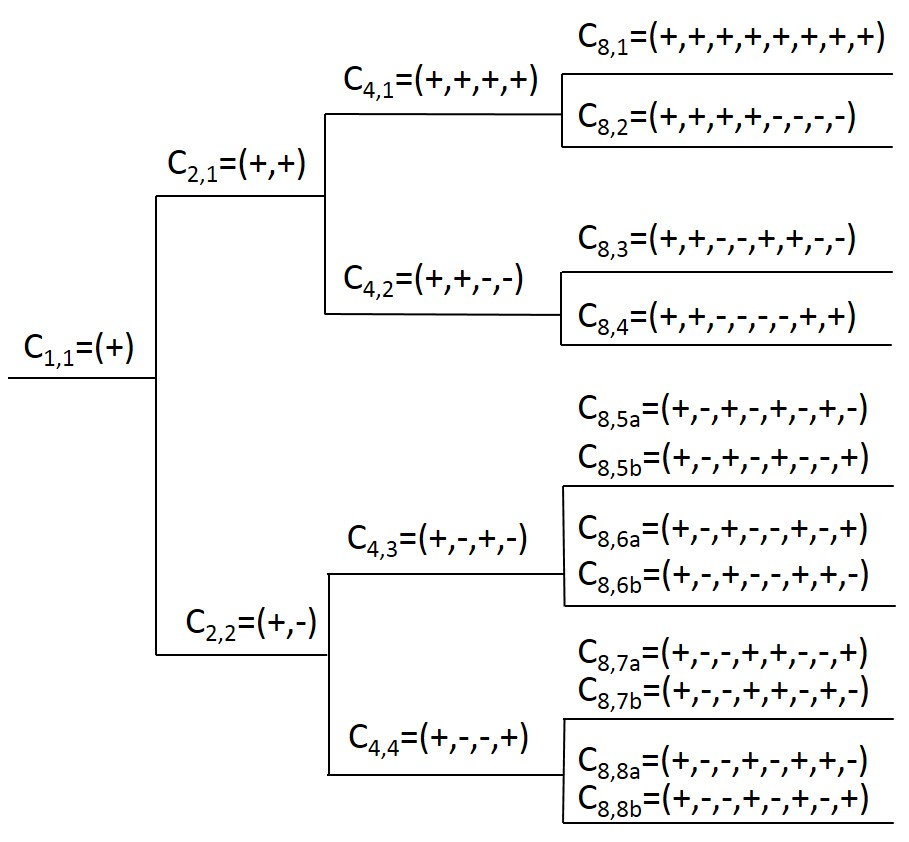}
	\caption{Partly Overloaded Variable Spreading Factor code tree}
	\label{fig:POVSF}
\end{figure}

As we already stated, in a second step, we now would like to combine the proposed partly overloaded spreading sequences with the idea of a variable spreading factor, which leads to a maximum of flexibility. To do so, we arrange the sequences described in the preceding section on a code tree, as it is known with standard OVSF codes (cf. figure \ref{fig:POVSF}). This leads us to a code tree, which has the following properties:
\begin{itemize}
\item All spreading sequences of the upper half of the code tree are mutually orthogonal to every other sequence in the same layer and to every other code in the tree, as long as one sequence is not a mother code of the second sequence.
\item In the lower half of the code tree, overloading is occurring as soon as more than one code per branch and layer is assigned. If not, the code tree is equivalent to a standard OVSF code tree.
\item Depending on our constraints concerning latency, throughput and bit error rate, when can now choose, whether to invest resources in a larger spreading factor $SF$, in a lower code rate or in a more robust modulation order. 
\end{itemize}
\section{Simulation}
As already stated, we have several parameters which affect the system performance w.r.t. BER, PER and throughput:
\begin{itemize}
\item the spreading factor (SF)
\item the signal to noise ratio (SNR)
\item the number of users ($N_U$), which effects the amount of overloading in the system
\item the sending probability of every user ($P_S$), which is also related to the overloading
\item the FEC code rate
\item the modulation order.
\end{itemize}
In order to enable a numerical evaluation the presented spreading sequences, a simulation has been performed which will be described in the following.
\subsection{Setup}

We simulated a setup with the parameters shown in table \ref{tab:parameters}. Please note, that these parameters can vary in some simulations, so that the values stated in table \ref{tab:parameters} hold true if not otherwise stated.
\begin{table}%[h]
\caption{Simulation Setup}
\centering
\begin{tabular}{l|l|l}
\textbf{Parameter} 						 & \textbf{Machine-type}  & \textbf{Best-Effort}\\
\hline
Number of users								 & $N_{MT}=4$							& $N_{BE}=6$\\
Number of iterations&\multicolumn{2}{c}{}\\ per simulation&\multicolumn{2}{c}{10.000}\\
Sending probability						 & \multicolumn{2}{c}{$P_S = 0.5$}\\
Packet length									 & \multicolumn{2}{c}{128 Bit}\\
Spreading Factor							 & \multicolumn{2}{c}{$SF=8$}\\
Modulation										 & \multicolumn{2}{c}{QPSK}\\
FEC code rate									 & \multicolumn{2}{c}{1}\\
Signal to noise ratio (SNR)    & \multicolumn{2}{c}{10 dB}\\
Synchronization								 & \multicolumn{2}{c}{perfect}\\
Channel Estimation 						 & \multicolumn{2}{c}{perfect}\\
\end{tabular}
\label{tab:parameters}
\end{table}

The signal processing chain consists of the following components: a data source, the digital modulation, the spreading, an AWGN channel, the despreading, the demodulation and a data sink. These components will now be described in detail.
\subsubsection{Data Source}
In our simulation, we use a simple packet source, which is generating packets with a length of 128 Bit and with a probability of $P_S$.

\subsubsection{Digital Modulation}
We use QPSK modulation where not stated otherwise.

\subsubsection{Spreading}
The modulated user data vectors $A_i$ are then spread using the spreading code vectors $C_i$ by applying the Kronecker product, yielding to the spread signal $S_i$:
\begin{equation}
S_i = A_i \otimes C_i.
\end{equation}

\subsubsection{Channel Model}
Since in this paper, we focus on the performance of the proposed spreading sequences, we assume, that channel distortions have been equalized by some preceding processing steps (e.g. when CDMA is used within a MC-CDMA system) and that only AWGN is applied to the CDMA part of the system. The spreaded user signals $S_i$ are however superposed yielding to the signal $S$:
\begin{equation}
S = \sum_{i=1}^{N_U} S_i.
\end{equation}

\subsubsection{Despreading}
The despreading of the spreaded signal $S$ is done by multiplying the repeated code vectors $\hat{C_i} = (C_i, C_i, ..., C_i)$ for $N_B$ times component-wise with $S$, yielding to the despread user signal $\tilde{D_i}$:
\begin{equation}
\tilde{D_i} = \hat{C_i} \odot S, 
\end{equation}
with $\odot$ being the Hadamard product.
Finally the despread user signal for the $i$th user $D_i$ is achieved by applying ``integrate and dump'' to $\tilde{D_i}$:
\begin{equation}
D_i = (d_{i,1}, ..., d_{i,N_B})
\end{equation}
with
\begin{equation}
d_{i,j} = \sum_{x=(j-1) \cdot SF + 1}^{j \cdot SF} \tilde{d_{i,x}} / SF.
\end{equation}

\subsubsection{Data Sink}
In the data sink, the Bit Error Rate ($BER$) is calculated. BER over all $N_I$ iterations and all users $N_U$ is calculated as
\begin{equation}
BER = \frac{\sum_{a=1}^{N_I} \sum_{i=1}^{N_U} \sum_{j=1}^{N_B} \|d_{i,j}^a - u_{i,j}^a\|}{N_I \cdot N_U \cdot N_B}
\end{equation}
with $d_{i,j}^a$ being the $j$th received bit of the $a$th packet of user $i$.

\subsection{Results}
Next, some simulation results are presented. To this purpose, the system performance is investigated as a function of the SNR, number of users, and sending probability. In a second step, the trade-off between FEC code rate, spreading factor and modulation order is investigated.

\subsubsection{Variation of the SNR}
\begin{figure}%[h!]
	%\centering
		% This file was created by matlab2tikz v0.4.3.
% Copyright (c) 2008--2013, Nico Schlömer <nico.schloemer@gmail.com>
% All rights reserved.
% 
% The latest updates can be retrieved from
%   http://www.mathworks.com/matlabcentral/fileexchange/22022-matlab2tikz
% where you can also make suggestions and rate matlab2tikz.
% 
%
% defining custom colors
\definecolor{mycolor1}{rgb}{0,0.447,0.741}%
\definecolor{mycolor2}{rgb}{0.85,0.325,0.098}%
\definecolor{mycolor3}{rgb}{0.929,0.694,0.125}%
\definecolor{mycolor4}{rgb}{0.494,0.184,0.556}%%
\begin{tikzpicture}

\begin{axis}[%
width=2.5in,
height=2.0in,
scale only axis,
separate axis lines,
every outer x axis line/.append style={darkgray!60!black},
every x tick label/.append style={font=\color{darkgray!60!black}},
xmin=0,
xmax=20,
xlabel={SNR in dB},
xmajorgrids,
every outer y axis line/.append style={darkgray!60!black},
every y tick label/.append style={font=\color{darkgray!60!black}},
ymode=log,
ymin=1e-7,%1e-35,
ymax=1,
yminorticks=true,
ylabel={Bit Error Rate},
ymajorgrids,
yminorgrids,
legend style={at={(0.784705438346378,0.83806076300785)},anchor=south west,draw=darkgray!60!black,fill=white,legend cell align=left, font=\tiny},
label style={font=\scriptsize},
tick label style={font=\scriptsize},
legend pos=south west
]
\addplot [
color=mycolor1,
solid,
line width=1.0pt,
mark=diamond
]
table[row sep=crcr]{
1 0.0463716796875\\
2 0.0347243637304687\\
3 0.024879058373873\\
4 0.0164265113433374\\
5 0.00996648640113433\\
6 0.00550685602364011\\
7 0.00277437881060236\\
8 0.00116648837538106\\
9 0.000447772898837538\\
10 0.000123091652289884\\
11 2.6184184165229e-05\\
12 5.47136841841652e-06\\
13 1.36773463684184e-06\\
14 1.36773463684184e-10\\
15 1.36773463684184e-14\\
16 1.36773463684184e-18\\
17 1.36773463684184e-22\\
18 1.36773463684184e-26\\
19 1.36773463684184e-30\\
20 1.36773463684184e-34\\
};
\addlegendentry{Machine-type traffic};
\addplot [
color=mycolor2,
solid,
line width=1.0pt,
mark=asterisk
]
table[row sep=crcr]{
1 0.0729346354166667\\
2 0.064802345546875\\
3 0.0582969750262214\\
4 0.0509491890725026\\
5 0.0451353032522406\\
6 0.0413823781136586\\
7 0.037302836154478\\
8 0.0334190948669488\\
9 0.0315228731594867\\
10 0.0290873970789826\\
11 0.0281047316563746\\
12 0.0261739042231656\\
13 0.0256261850987557\\
14 0.0259775626185099\\
15 0.0259436133812618\\
16 0.0261688704030048\\
17 0.0254332158453736\\
18 0.0258547568632512\\
19 0.0255775854756863\\
20 0.0257792504668809\\
};
\addlegendentry{Best-effort traffic};
%\addplot [
%color=mycolor3,
%dashed,
%line width=1.0pt,
%mark=cube,
%mark options={scale=1,solid}
%]
%table[row sep=crcr]{
%1 0.0376893554687500\\
%2 0.0269011322167969\\
%3 0.0181445846444717\\
%4 0.0116347246147144\\
%5 0.00647587050371147\\
%6 0.00327008118080037\\
%7 0.00147962388311808\\
%8 0.000544972181138312\\
%9 0.000176616997218114\\
%10 4.48418804497218e-05\\
%11 7.81698418804497e-06\\
%12 6.84375448418805e-07\\
%13 6.84375448418804e-11\\
%14 6.84375448418805e-15\\
%15 6.84375448418805e-19\\
%16 6.84375448418805e-23\\
%17 6.84375448418805e-27\\
%18 6.84375448418805e-31\\
%19 6.84375448418805e-35\\
%20 6.84375448418805e-39\\
%};
%\addlegendentry{Hadamard $P_S = 0.5$};
\addplot [
color=mycolor4,
dashed,
line width=1.0pt,
mark=star,
mark options={scale=1,solid}
]
table[row sep=crcr]{
1 0.0562675781250000\\
2 0.0424515251953125\\
3 0.0296156709337695\\
4 0.0194068678170934\\
5 0.0115276242805317\\
6 0.00612556682492805\\
7 0.00291604224418249\\
8 0.00119814316672442\\
9 0.000405686220566672\\
10 0.000105997599872057\\
11 2.07137247599872e-05\\
12 2.44347762247600e-06\\
13 1.95556847762248e-07\\
14 1.95556847762248e-11\\
15 1.95556847762248e-15\\
16 1.95556847762248e-19\\
17 1.95556847762248e-23\\
18 1.95556847762248e-27\\
19 1.95556847762248e-31\\
20 1.95556847762248e-35\\
};
\addlegendentry{Hadamard $\widehat{P_S} = 0.625$};
\end{axis}
\end{tikzpicture}%
	\caption{BER vs. SNR}
	\label{fig:BERvsSNR}
\end{figure}
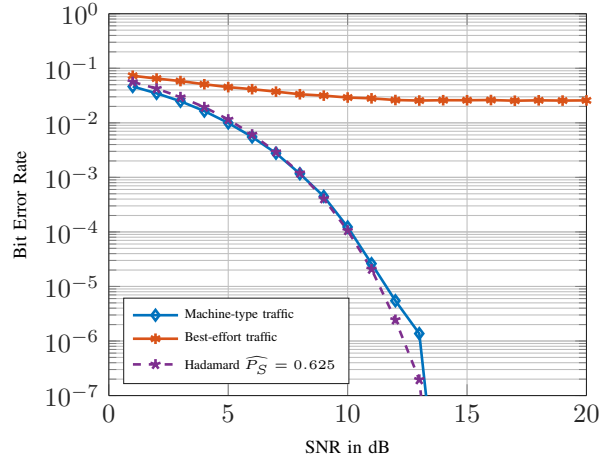

In a first simulation, we would like to investigate the system performance for different Signal-to-noise ratios (SNRs). Therefore, the SNR is varied from $0dB$ to $20dB$ in steps of $1dB$. The result of this simulation is depicted in figure \ref{fig:BERvsSNR}.
As we can observe, the bit error rate (BER) of best-effort traffic in this scenario ($N_U=10$ and $P_S=0.5$, i.e. overloading occurs) is hardly improving with better SNR, whereas the BER of machine-type traffic improves dramatically. This shows, that orthogonality of machine-type users and their spreading sequences can be maintained while overloading best-effort users. The dashed curve shows the BER performance of a pure Hadamard set with $SF=8$ for the sake of comparison. Since a Hadamard set of spreading sequences with $SF=8$ is able to support up to $N_H=8$ different users, the sending probability $\widehat{P_S}$ has been set to
\begin{equation}
\widehat{P_S} = \frac{N_U}{N_H} \times P_S=\frac{10}{8} \times 0.5 = 0.625
\end{equation} to ensure a realistic comparison. As we can observe, the BER performance of this pure Hadamard set with $SF=8$ is identical to the performance of our proposed sequences for the case of machine-type traffic.

\subsubsection{Variation of the sending probability}
\begin{figure}%[h!]
	%\centering
		%\input{figures/BERvsPS}
		% This file was created by matlab2tikz v0.4.3.
% Copyright (c) 2008--2013, Nico Schlömer <nico.schloemer@gmail.com>
% All rights reserved.
% 
% The latest updates can be retrieved from
%   http://www.mathworks.com/matlabcentral/fileexchange/22022-matlab2tikz
% where you can also make suggestions and rate matlab2tikz.
% 
%
% defining custom colors
\definecolor{mycolor1}{rgb}{0,0.447,0.741}%
\definecolor{mycolor2}{rgb}{0.85,0.325,0.098}%
\definecolor{mycolor3}{rgb}{0.929,0.694,0.125}%
\begin{tikzpicture}

\begin{axis}[%
width=2.5in,
height=2.0in,
scale only axis,
separate axis lines,
every outer x axis line/.append style={darkgray!60!black},
every x tick label/.append style={font=\color{darkgray!60!black}},
xmin=0.1,
xmax=1,
xlabel={Sending Probability},
xmajorgrids,
every outer y axis line/.append style={darkgray!60!black},
every y tick label/.append style={font=\color{darkgray!60!black}},
ymode=log,
ymin=1e-06,
ymax=1,
yminorticks=true,
ylabel={Bit Error Rate},
ymajorgrids,
yminorgrids,
legend style={draw=darkgray!60!black,fill=white,legend cell align=left, font=\tiny},
label style={font=\scriptsize},
tick label style={font=\scriptsize},
legend pos=south east
]
\addplot [
color=mycolor1,
solid,
line width=1.0pt,
mark=diamond
]
table[row sep=crcr]{
0 0\\
0.1 0\\
0.2 2.34375e-06\\
0.3 1.3281484375e-05\\
0.4 4.35560156484375e-05\\
0.5 0.000116019980601565\\
0.6 0.00026251160199806\\
0.7 0.0005203387511602\\
0.8 0.000929739533875116\\
0.9 0.00152958516145339\\
1 0.00233511389601615\\
};
\addlegendentry{Machine-type traffic};
\addplot [
color=mycolor2,
solid,
line width=1.0pt,
mark=asterisk
]
table[row sep=crcr]{
0 0\\
0.1 0.000394270833333333\\
0.2 0.00271605505208333\\
0.3 0.00829493306383854\\
0.4 0.0168672357433064\\
0.5 0.0291231710985743\\
0.6 0.0452560373171099\\
0.7 0.065363640187065\\
0.8 0.0889786717806854\\
0.9 0.115900824950511\\
1 0.146768230707495\\
};
\addlegendentry{Best-effort traffic};
%\addplot [
%color=mycolor3,
%dashed,
%line width=1.0pt,
%mark=cube,
%mark options={scale=1,solid}
%]
%table[row sep=crcr]{
%0 0\\
%0.1 0\\
%0.2 1.75781250000000e-06\\
%0.3 5.46892578125000e-06\\
%0.4 1.76763281425781e-05\\
%0.5 4.45330176328143e-05\\
%0.6 8.40864845517633e-05\\
%0.7 0.000171199814898455\\
%0.8 0.000304411651231490\\
%0.9 0.000505206222415123\\
%1 0.000791945051872242\\
%};
%\addlegendentry{Fully loaded Hadamard set};
\end{axis}
\end{tikzpicture}%
	\caption{BER vs. Sending Probability}
	\label{fig:BERvsPS}
\end{figure}
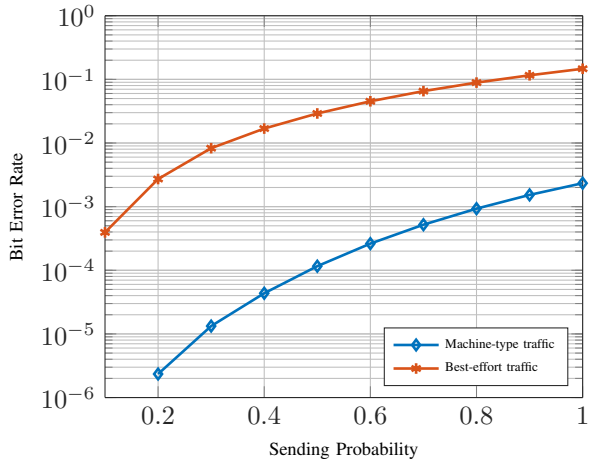

In a second step, we are intending to investigate the impact of the sending probability on the system performance. For this purpose, $P_S$ is varied from $0\%$ to $100\%$ in steps of $10\%$ and the result is depicted in figure \ref{fig:BERvsPS}.
As one would expect, the Bit Error Rate (BER) for both machine-type as well as best-effort traffic worsens with a growing sending probability. Due to its orthogonality, machine-type traffic has a better BER with about two to three orders of magnitude difference.

\subsubsection{Variation of the number of users}
\begin{figure}%[h!]
	%\centering
		% This file was created by matlab2tikz v0.4.3.
% Copyright (c) 2008--2013, Nico Schlömer <nico.schloemer@gmail.com>
% All rights reserved.
% 
% The latest updates can be retrieved from
%   http://www.mathworks.com/matlabcentral/fileexchange/22022-matlab2tikz
% where you can also make suggestions and rate matlab2tikz.
% 
\begin{tikzpicture}

\begin{axis}[%
width=2.5in,
height=2.0in,
view={-37.5}{30},
scale only axis,
every outer x axis line/.append style={darkgray!60!black},
every x tick label/.append style={font=\color{darkgray!60!black}},
xmin=1,
xmax=8,
xlabel={$N_{BE}$},
xmajorgrids,
xtick={1,...,8},
every outer y axis line/.append style={darkgray!60!black},
every y tick label/.append style={font=\color{darkgray!60!black}},
ymin=1,
ymax=4,
ylabel={$N_{RT}$},
ymajorgrids,
ytick={1,...,4},
every outer z axis line/.append style={darkgray!60!black},
every z tick label/.append style={font=\color{darkgray!60!black}},
zmin=0,
zmax=0.0003,
zlabel={BER},
zmajorgrids,
axis x line*=bottom,
axis y line*=left,
axis z line*=left,
label style={font=\scriptsize},
tick label style={font=\scriptsize}
]

\addplot3[%
surf,
shader=faceted,
draw=black,
colormap={mymap}{[1pt] rgb(0pt)=(0.2081,0.1663,0.5292); rgb(1pt)=(0.211624,0.189781,0.577676); rgb(2pt)=(0.212252,0.213771,0.626971); rgb(3pt)=(0.2081,0.2386,0.677086); rgb(4pt)=(0.195905,0.264457,0.7279); rgb(5pt)=(0.170729,0.291938,0.779248); rgb(6pt)=(0.125271,0.324243,0.830271); rgb(7pt)=(0.0591333,0.359833,0.868333); rgb(8pt)=(0.0116952,0.38751,0.881957); rgb(9pt)=(0.00595714,0.408614,0.882843); rgb(10pt)=(0.0165143,0.4266,0.878633); rgb(11pt)=(0.0328524,0.443043,0.871957); rgb(12pt)=(0.0498143,0.458571,0.864057); rgb(13pt)=(0.0629333,0.47369,0.855438); rgb(14pt)=(0.0722667,0.488667,0.8467); rgb(15pt)=(0.0779429,0.503986,0.838371); rgb(16pt)=(0.0793476,0.520024,0.831181); rgb(17pt)=(0.0749429,0.537543,0.826271); rgb(18pt)=(0.0640571,0.556986,0.823957); rgb(19pt)=(0.0487714,0.577224,0.822829); rgb(20pt)=(0.0343429,0.596581,0.819852); rgb(21pt)=(0.0265,0.6137,0.8135); rgb(22pt)=(0.0238905,0.628662,0.803762); rgb(23pt)=(0.0230905,0.641786,0.791267); rgb(24pt)=(0.0227714,0.653486,0.776757); rgb(25pt)=(0.0266619,0.664195,0.760719); rgb(26pt)=(0.0383714,0.674271,0.743552); rgb(27pt)=(0.0589714,0.683757,0.725386); rgb(28pt)=(0.0843,0.692833,0.706167); rgb(29pt)=(0.113295,0.7015,0.685857); rgb(30pt)=(0.145271,0.709757,0.664629); rgb(31pt)=(0.180133,0.717657,0.642433); rgb(32pt)=(0.217829,0.725043,0.619262); rgb(33pt)=(0.258643,0.731714,0.595429); rgb(34pt)=(0.302171,0.737605,0.571186); rgb(35pt)=(0.348167,0.742433,0.547267); rgb(36pt)=(0.395257,0.7459,0.524443); rgb(37pt)=(0.44201,0.748081,0.503314); rgb(38pt)=(0.487124,0.749062,0.483976); rgb(39pt)=(0.530029,0.749114,0.466114); rgb(40pt)=(0.570857,0.748519,0.44939); rgb(41pt)=(0.609852,0.747314,0.433686); rgb(42pt)=(0.6473,0.7456,0.4188); rgb(43pt)=(0.683419,0.743476,0.404433); rgb(44pt)=(0.71841,0.741133,0.390476); rgb(45pt)=(0.752486,0.7384,0.376814); rgb(46pt)=(0.785843,0.735567,0.363271); rgb(47pt)=(0.818505,0.732733,0.34979); rgb(48pt)=(0.850657,0.7299,0.336029); rgb(49pt)=(0.882433,0.727433,0.3217); rgb(50pt)=(0.913933,0.725786,0.306276); rgb(51pt)=(0.944957,0.726114,0.288643); rgb(52pt)=(0.973895,0.731395,0.266648); rgb(53pt)=(0.993771,0.745457,0.240348); rgb(54pt)=(0.999043,0.765314,0.216414); rgb(55pt)=(0.995533,0.786057,0.196652); rgb(56pt)=(0.988,0.8066,0.179367); rgb(57pt)=(0.978857,0.827143,0.163314); rgb(58pt)=(0.9697,0.848138,0.147452); rgb(59pt)=(0.962586,0.870514,0.1309); rgb(60pt)=(0.958871,0.8949,0.113243); rgb(61pt)=(0.959824,0.921833,0.0948381); rgb(62pt)=(0.9661,0.951443,0.0755333); rgb(63pt)=(0.9763,0.9831,0.0538)},
mesh/rows=8]
table[row sep=crcr,header=false] {
1 1 0\\
1 2 2.61734375585965e-09\\
1 3 5.28125398457293e-07\\
1 4 1.7720906797418e-06\\
2 1 0\\
2 2 3.90625261734376e-07\\
2 3 2.08338614587318e-06\\
2 4 6.83611470906797e-06\\
3 1 0\\
3 2 1.17191406252617e-06\\
3 3 7.55229167194792e-06\\
3 4 2.22663086114709e-05\\
4 1 1.5625e-06\\
4 2 5.07824219140625e-06\\
4 3 2.10945052291672e-05\\
4 4 4.06272266308611e-05\\
5 1 5.46890625e-06\\
5 2 1.87505078242191e-05\\
5 3 3.93250261171896e-05\\
5 4 6.97306252226631e-05\\
6 1 1.1719296890625e-05\\
6 2 2.96893750507824e-05\\
6 3 7.23997658359451e-05\\
6 4 0.000117975723062522\\
7 1 3.12511719296891e-05\\
7 2 5.97685939375051e-05\\
7 3 0.000111465573309917\\
7 4 0.000181652422572306\\
8 1 5.2346875117193e-05\\
8 2 0.000109380976859394\\
8 3 0.000190375729890664\\
8 4 0.000271697852742257\\
};
\end{axis}
\end{tikzpicture}%
	\caption{BER of machine-type traffic users vs. number of users}
	\label{fig:BERvsNU_RT}
\end{figure}
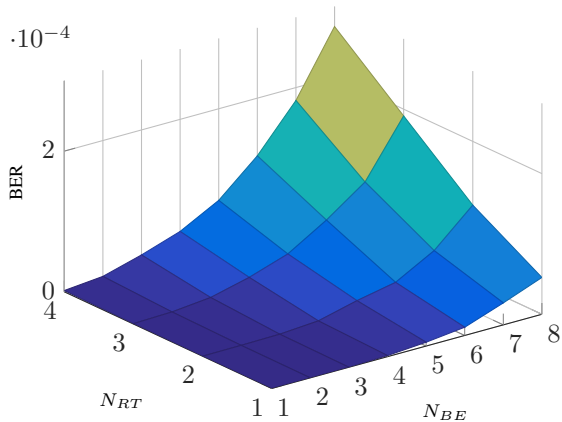

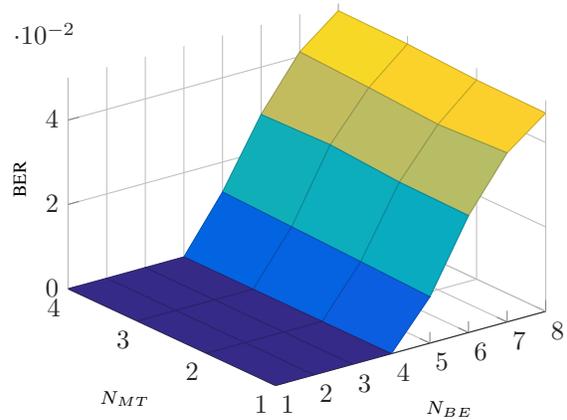
\begin{figure}%[h!]
	%\centering
		% This file was created by matlab2tikz v0.4.3.
% Copyright (c) 2008--2013, Nico Schlömer <nico.schloemer@gmail.com>
% All rights reserved.
% 
% The latest updates can be retrieved from
%   http://www.mathworks.com/matlabcentral/fileexchange/22022-matlab2tikz
% where you can also make suggestions and rate matlab2tikz.
% 
\begin{tikzpicture}

\begin{axis}[%
width=2.5in,
height=2.0in,
view={-37.5}{30},
scale only axis,
every outer x axis line/.append style={darkgray!60!black},
every x tick label/.append style={font=\color{darkgray!60!black}},
xmin=1,
xmax=8,
xlabel={$N_{BE}$},
xmajorgrids,
xtick={1,...,8},
every outer y axis line/.append style={darkgray!60!black},
every y tick label/.append style={font=\color{darkgray!60!black}},
ymin=1,
ymax=4,
ylabel={$N_{MT}$},
ymajorgrids,
ytick={1,...,4},
every outer z axis line/.append style={darkgray!60!black},
every z tick label/.append style={font=\color{darkgray!60!black}},
zmin=0,
zmax=0.05,
zlabel={BER},
zmajorgrids,
axis x line*=bottom,
axis y line*=left,
axis z line*=left,
label style={font=\scriptsize},
tick label style={font=\scriptsize}
]

\addplot3[%
surf,
shader=faceted,
draw=black,
colormap={mymap}{[1pt] rgb(0pt)=(0.2081,0.1663,0.5292); rgb(1pt)=(0.211624,0.189781,0.577676); rgb(2pt)=(0.212252,0.213771,0.626971); rgb(3pt)=(0.2081,0.2386,0.677086); rgb(4pt)=(0.195905,0.264457,0.7279); rgb(5pt)=(0.170729,0.291938,0.779248); rgb(6pt)=(0.125271,0.324243,0.830271); rgb(7pt)=(0.0591333,0.359833,0.868333); rgb(8pt)=(0.0116952,0.38751,0.881957); rgb(9pt)=(0.00595714,0.408614,0.882843); rgb(10pt)=(0.0165143,0.4266,0.878633); rgb(11pt)=(0.0328524,0.443043,0.871957); rgb(12pt)=(0.0498143,0.458571,0.864057); rgb(13pt)=(0.0629333,0.47369,0.855438); rgb(14pt)=(0.0722667,0.488667,0.8467); rgb(15pt)=(0.0779429,0.503986,0.838371); rgb(16pt)=(0.0793476,0.520024,0.831181); rgb(17pt)=(0.0749429,0.537543,0.826271); rgb(18pt)=(0.0640571,0.556986,0.823957); rgb(19pt)=(0.0487714,0.577224,0.822829); rgb(20pt)=(0.0343429,0.596581,0.819852); rgb(21pt)=(0.0265,0.6137,0.8135); rgb(22pt)=(0.0238905,0.628662,0.803762); rgb(23pt)=(0.0230905,0.641786,0.791267); rgb(24pt)=(0.0227714,0.653486,0.776757); rgb(25pt)=(0.0266619,0.664195,0.760719); rgb(26pt)=(0.0383714,0.674271,0.743552); rgb(27pt)=(0.0589714,0.683757,0.725386); rgb(28pt)=(0.0843,0.692833,0.706167); rgb(29pt)=(0.113295,0.7015,0.685857); rgb(30pt)=(0.145271,0.709757,0.664629); rgb(31pt)=(0.180133,0.717657,0.642433); rgb(32pt)=(0.217829,0.725043,0.619262); rgb(33pt)=(0.258643,0.731714,0.595429); rgb(34pt)=(0.302171,0.737605,0.571186); rgb(35pt)=(0.348167,0.742433,0.547267); rgb(36pt)=(0.395257,0.7459,0.524443); rgb(37pt)=(0.44201,0.748081,0.503314); rgb(38pt)=(0.487124,0.749062,0.483976); rgb(39pt)=(0.530029,0.749114,0.466114); rgb(40pt)=(0.570857,0.748519,0.44939); rgb(41pt)=(0.609852,0.747314,0.433686); rgb(42pt)=(0.6473,0.7456,0.4188); rgb(43pt)=(0.683419,0.743476,0.404433); rgb(44pt)=(0.71841,0.741133,0.390476); rgb(45pt)=(0.752486,0.7384,0.376814); rgb(46pt)=(0.785843,0.735567,0.363271); rgb(47pt)=(0.818505,0.732733,0.34979); rgb(48pt)=(0.850657,0.7299,0.336029); rgb(49pt)=(0.882433,0.727433,0.3217); rgb(50pt)=(0.913933,0.725786,0.306276); rgb(51pt)=(0.944957,0.726114,0.288643); rgb(52pt)=(0.973895,0.731395,0.266648); rgb(53pt)=(0.993771,0.745457,0.240348); rgb(54pt)=(0.999043,0.765314,0.216414); rgb(55pt)=(0.995533,0.786057,0.196652); rgb(56pt)=(0.988,0.8066,0.179367); rgb(57pt)=(0.978857,0.827143,0.163314); rgb(58pt)=(0.9697,0.848138,0.147452); rgb(59pt)=(0.962586,0.870514,0.1309); rgb(60pt)=(0.958871,0.8949,0.113243); rgb(61pt)=(0.959824,0.921833,0.0948381); rgb(62pt)=(0.9661,0.951443,0.0755333); rgb(63pt)=(0.9763,0.9831,0.0538)},
mesh/rows=8]
table[row sep=crcr,header=false] {
1 1 0\\
1 2 4.75370768243382e-06\\
1 3 4.77620153417841e-06\\
1 4 7.90831378553865e-06\\
2 1 0\\
2 2 3.9109551497877e-07\\
2 3 7.81718904741672e-07\\
2 4 6.6412561087305e-06\\
3 1 5.20833333333333e-07\\
3 2 2.86460942192896e-06\\
3 3 1.01563021303612e-05\\
3 4 2.0312942766598e-05\\
4 1 3.3203515625e-06\\
4 2 8.78927734570782e-06\\
4 3 2.2461699222661e-05\\
4 4 3.67202734707087e-05\\
5 1 0.0109701565156281\\
5 2 0.0116596882031422\\
5 3 0.0123014080469359\\
5 4 0.0129262529376219\\
6 1 0.0277134141797096\\
6 2 0.0280043570573503\\
6 3 0.0292199053256706\\
6 4 0.0287762073960781\\
7 1 0.0400599647212154\\
7 2 0.0391608825163192\\
7 3 0.0402562992061708\\
7 4 0.0410909709963482\\
8 1 0.0469833880594131\\
8 2 0.0471876062647202\\
8 3 0.0480692450824305\\
8 4 0.0483417790537122\\
};
\end{axis}
\end{tikzpicture}%
	\caption{BER of best-effort traffic users vs. number of users}
	\label{fig:BERvsNU_NRT}
\end{figure}

Furthermore, we investigated the influence of the number of users in the system, both for best-effort type users as well as for machine-type traffic users. The results are depicted in figure \ref{fig:BERvsNU_RT} for the BER of the machine-type users and in figure \ref{fig:BERvsNU_NRT} for best-effort traffic users.

As we can see, the BER in both cases worsens, as soon as overloading occurs, i.e. $N_{BE} > 4$. Due to the orthogonality in the case of machine-type traffic, this group of users experiences a better BER with at least two orders of magnitude compared to the users with best-effort traffic. 

\subsubsection{Investigation of the trade-off between FEC code rate and spreading factor}
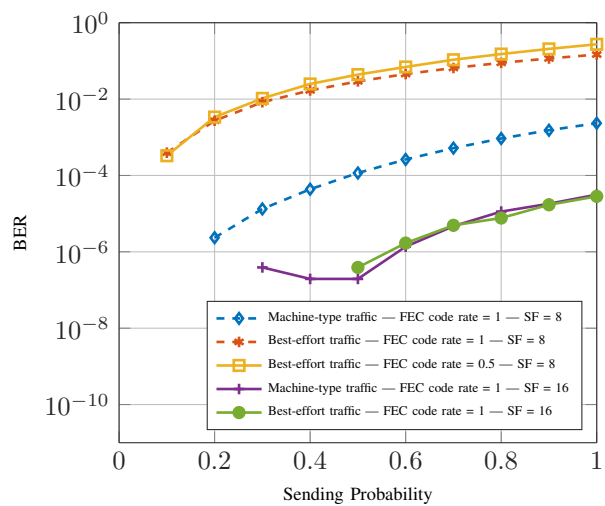
\begin{figure}%[h!]
	%\centering
		% This file was created by matlab2tikz v0.4.3.
% Copyright (c) 2008--2013, Nico Schlömer <nico.schloemer@gmail.com>
% All rights reserved.
% 
% The latest updates can be retrieved from
%   http://www.mathworks.com/matlabcentral/fileexchange/22022-matlab2tikz
% where you can also make suggestions and rate matlab2tikz.
% 
%
% defining custom colors
\definecolor{mycolor1}{rgb}{0,0.447,0.741}%
\definecolor{mycolor2}{rgb}{0.85,0.325,0.098}%
\definecolor{mycolor3}{rgb}{0.929,0.694,0.125}%
\definecolor{mycolor4}{rgb}{0.494,0.184,0.556}%
\definecolor{mycolor5}{rgb}{0.466,0.674,0.188}%
\definecolor{mycolor6}{rgb}{0.301,0.745,0.933}%
\begin{tikzpicture}

\begin{axis}[%
width=2.5in,
height=2.2in,
scale only axis,
separate axis lines,
every outer x axis line/.append style={darkgray!60!black},
every x tick label/.append style={font=\color{darkgray!60!black}},
xmin=0,
xmax=1,
xlabel={Sending Probability},
xmajorgrids,
every outer y axis line/.append style={darkgray!60!black},
every y tick label/.append style={font=\color{darkgray!60!black}},
ymode=log,
ymin=1e-11,%0.0001,
ymax=1,
yminorticks=true,
ylabel={BER},
ymajorgrids,
yminorgrids,
legend style={draw=darkgray!60!black,fill=white,legend cell align=left, font=\tiny},
label style={font=\scriptsize},
tick label style={font=\scriptsize},
legend pos=south east
]
\addplot [
color=mycolor1,
dashed,
line width=1.0pt,
mark=diamond,
mark options={scale=1,solid}
]
table[row sep=crcr]{
0 0\\
0.1 0\\
0.2 2.34375e-06\\
0.3 1.3281484375e-05\\
0.4 4.35560156484375e-05\\
0.5 0.000116019980601565\\
0.6 0.00026251160199806\\
0.7 0.0005203387511602\\
0.8 0.000929739533875116\\
0.9 0.00152958516145339\\
1 0.00233511389601615\\
};
\addlegendentry{Machine-type traffic | FEC code rate = 1 | SF = 8};
\addplot [
color=mycolor2,
dashed,
line width=1.0pt,
mark=asterisk,
mark options={scale=1,solid}
]
table[row sep=crcr]{
0 0\\
0.1 0.000394270833333333\\
0.2 0.00271605505208333\\
0.3 0.00829493306383854\\
0.4 0.0168672357433064\\
0.5 0.0291231710985743\\
0.6 0.0452560373171099\\
0.7 0.065363640187065\\
0.8 0.0889786717806854\\
0.9 0.115900824950511\\
1 0.146768230707495\\
};
\addlegendentry{Best-effort traffic | FEC code rate = 1 | SF = 8};
\addplot [
color=mycolor3,
solid,
line width=1.0pt,
mark=square,
mark options={scale=1,solid}
]
table[row sep=crcr]{
0 0\\
0.1 0.000328255208333333\\
0.2 0.0033827151171875\\
0.3 0.0104589320215117\\
0.4 0.0249300823515355\\
0.5 0.0437954617582352\\
0.6 0.0693922701711758\\
0.7 0.107515923602017\\
0.8 0.151196038050694\\
0.9 0.206617593562138\\
1 0.272745661759356\\
};
\addlegendentry{Best-effort traffic | FEC code rate = 0.5 | SF = 8};
\addplot [
color=mycolor4,
solid,
line width=1.0pt,
mark=+,
mark options={scale=1,solid}
]
table[row sep=crcr]{
0 0\\
0.1 0\\
0.2 0\\
0.3 3.90625e-07\\
0.4 1.953515625e-07\\
0.5 1.9533203515625e-07\\
0.6 1.36720703320352e-06\\
0.7 4.68763672070332e-06\\
0.8 1.13285937636721e-05\\
0.9 1.81651953593764e-05\\
1 3.14471290195359e-05\\
};
\addlegendentry{Machine-type traffic | FEC code rate = 1 | SF = 16};
\addplot [
color=mycolor5,
solid,
line width=1.0pt,
mark=*,
mark options={scale=1,solid}
]
table[row sep=crcr]{
0 0\\
0.1 0\\
0.2 0\\
0.3 0\\
0.4 0\\
0.5 3.90625e-07\\
0.6 1.69274739583333e-06\\
0.7 4.94808594140625e-06\\
0.8 7.68278647526081e-06\\
0.9 1.70580599453142e-05\\
1 2.83871224726612e-05\\
};
\addlegendentry{Best-effort traffic | FEC code rate = 1 | SF = 16};
\end{axis}
\end{tikzpicture}%
	\caption{BER vs Sending Probability for different values of SF and FEC code rate}
	\label{fig:BERvsFECvsSF}
\end{figure}

In a last simulation, we investigated the the trade-off between FEC code rate and spreading factor. In order to do so, we calculated the BER vs. the sending probability of ten users. The results are depicted in figure \ref{fig:BERvsFECvsSF}. We observe, that for best-effort traffic, investing in a higher spreading factor returns with lower BER with four orders of magnitudes. This is because in this case of $SF=16$ the system is not overloaded and orthogonality is maintained for the best-effort type users as well. Applying a convolutional code with code rate $1/2$ brings no benefit in terms of BER for best-effort users, as we can see when comparing to the orange dashed line, which is equivalent to the illustration in figure \ref{fig:BERvsPS} and was included for the sake of comparison.
For machine-type traffic, an increased spreading factor of $SF=16$ results in a BER gain of two orders of magnitude compared to the overloaded case, which is represented by the blue dashed line. Since in the case of $SF=16$ the user spreading sequences are equivalent to standard Hadamard sequences, the performance w.r.t. to BER for machine-type as well as best-effort type users is identical.
The application of the FEC with code rate $1/2$ to machine-type traffic resulted in no bit errors and is therefore not shown in figure \ref{fig:BERvsFECvsSF}. Since machine-type users usually have strict requirements concerning the BER, the application of FEC would be the better choice compared to an increased spreading factor in this case.

\section{Conclusion}
When thinking of wireless industrial communication systems, the challenge of efficiently combining machine-type traffic as well as best effort traffic with one MAC method is a crucial one. In this paper, we have presented a CDMA-based approach which looks promising in this context. As we have shown with the proposed CDMA MAC it is possible to overload parts of the system, enabling a higher number of users with lower requirements w.r.t. BER, while maintaining orthogonality for users with strict BER requirements. This means that with the proposed sequences, we are able to support Hadamard-equivalent performance regarding BER for one part of the users while allowing to increase the overall number of users, in particular users with less strict BER requirements and rather low sending probability, at the same time.  The proposed MAC is therefore particularly appropriate, if the user group with lower BER requirements has a low sending probability as for example in wireless sensor networks. 

\section{Future Work}
As we have shown in this paper, the proposed CDMA-based MAC is a promising candidate when the efficient combination of machine-type traffic as well as best effort traffic in the same system is required. In the future it would be preferable to investigate this using higher spreading factors. It is expected, that the advantages of the proposed MAC increase with higher spreading factors, mainly for two reasons:
\begin{enumerate}
\item For higher spreading factors, the additional interference per spreading sequence is relatively small.
\item With higher spreading factors, more additional spreading sequences, generated according to this paper, are available.
\end{enumerate}
It would be interesting to investigate scenarios with lots of users with only best-effort requirements and low sending probability.

\section*{Acknowledgment}
This work has been supported by the Federal Ministry of Education and Research of the Federal Republic of Germany (Foerderkennzeichen 16KIS0267, HiFlecs). The authors alone are responsible for the content of the paper.

% Literaturverzeichnis
\bibliographystyle{IEEEtran}
\bibliography{references/references_v1,references/hybrid_mac,references/cdma}

% Generated by IEEEtran.bst, version: 1.13 (2008/09/30)
\begin{thebibliography}{1}
\providecommand{\url}[1]{#1}
\csname url@samestyle\endcsname
\providecommand{\newblock}{\relax}
\providecommand{\bibinfo}[2]{#2}
\providecommand{\BIBentrySTDinterwordspacing}{\spaceskip=0pt\relax}
\providecommand{\BIBentryALTinterwordstretchfactor}{4}
\providecommand{\BIBentryALTinterwordspacing}{\spaceskip=\fontdimen2\font plus
\BIBentryALTinterwordstretchfactor\fontdimen3\font minus
  \fontdimen4\font\relax}
\providecommand{\BIBforeignlanguage}[2]{{%
\expandafter\ifx\csname l@#1\endcsname\relax
\typeout{** WARNING: IEEEtran.bst: No hyphenation pattern has been}%
\typeout{** loaded for the language `#1'. Using the pattern for}%
\typeout{** the default language instead.}%
\else
\language=\csname l@#1\endcsname
\fi
#2}}
\providecommand{\BIBdecl}{\relax}
\BIBdecl

\bibitem{Nikaein2013}
N.~Nikaein, M.~Laner, K.~Zhou, P.~Svoboda, D.~Drajic, M.~Popovic, and S.~Krco,
  ``Simple traffic modeling framework for machine type communication,'' in
  \emph{Wireless Communication Systems (ISWCS 2013), Proceedings of the Tenth
  International Symposium on}.\hskip 1em plus 0.5em minus 0.4em\relax VDE,
  2013, pp. 1--5.

\bibitem{Sylvester1867}
J.~J. Sylvester, ``Lx. thoughts on inverse orthogonal matrices, simultaneous
  signsuccessions, and tessellated pavements in two or more colours, with
  applications to newton's rule, ornamental tile-work, and the theory of
  numbers,'' \emph{The London, Edinburgh, and Dublin Philosophical Magazine and
  Journal of Science}, vol.~34, no. 232, pp. 461--475, 1867.

\bibitem{Hadamard1893}
J.~{Hadamard}, ``\BIBforeignlanguage{French}{R\'esolution d'une question
  relative aux d\'eterminants.}'' \emph{\BIBforeignlanguage{French}{Bulletin
  des Sciences Math\'ematiques. Deuxi\`eme S\'erie}}, vol.~17, pp. 240--246,
  1893.

\bibitem{Adachi1997}
F.~Adachi, M.~Sawahashi, and K.~Okawa, ``Tree-structured generation of
  orthogonal spreading codes with different length for forward link of ds-cdma
  mobile radio,'' \emph{Electronic Letters}, vol.~33, no.~1, pp. 27--28,
  January 1997.

\bibitem{Hosseini2011}
\BIBentryALTinterwordspacing
S.~A. Hosseini, O.~Javidbakht, P.~Pad, and F.~Marvasti, ``A review on
  synchronous cdma systems: optimum overloaded codes, channel capacity, and
  power control,'' \emph{EURASIP Journal on Wireless Communications and
  Networking}, 2011. [Online]. Available:
  \url{http://jwcn.eurasipjournals.com/content/2011/1/62}
\BIBentrySTDinterwordspacing

\bibitem{Yang2000}
K.~Yang, Y.-K. Kim, and P.~V. Kumer, ``Quasi-orthogonal sequences for
  code-divison multiple-access systems,'' \emph{IEEE Transactions on
  Information Theory}, vol.~46, no.~3, pp. 982--992, May 2000.

\bibitem{Amadei2002}
M.~Amadei, U.~Manzoli, and M.~L. Merani, ``On the assignment of walsh and
  quasi-orthogonal codes in a multicarrier ds-cdma system with multiple classes
  of users,'' in \emph{Global Telecommunications Conference, 2002. GLOBECOM'02.
  IEEE}, vol.~1.\hskip 1em plus 0.5em minus 0.4em\relax IEEE, 2002, pp.
  841--845.

\bibitem{Kaiser1996}
S.~Kaiser, ``Trade-off between channel coding and spreading in multi-carrier
  cdma systems,'' in \emph{Spread Spectrum Techniques and Applications
  Proceedings, 1996., IEEE 4th International Symposium on}, vol.~3.\hskip 1em
  plus 0.5em minus 0.4em\relax IEEE, 1996, pp. 1366--1370.

\bibitem{Koukouvinos2008}
C.~Koukouvinos and S.~Stylianou, ``On skew-hadamard matrices,'' \emph{Discrete
  Mathematics}, vol. 308, no.~13, pp. 2723--2731, 2008.

\end{thebibliography}
%\bibliography{references/biblioOwnPubs,references/bibliography,references/biblioSolarMesh,references/rfc,references/SpectrumManagement}

\addtocounter{figure}{-1}\renewcommand{\thefigure}{\arabic{figure}a}
\end{document}